# Eigenmodal Analysis of Anderson Localization: Applications to Photonic Lattices and Bose-Einstein Condensates

Guanwen Ying, Guennadi Kouzaev*
Department of Electronics and Telecommunications
NTNU, Norwegian University of Science and Technology
O.S. Bragstads plass 2B, Trondheim 7491
Norway
guanwen.ying@iet.ntnu.no, guennadi.kouzaev@iet.ntnu.no*
Tel. +47 73596369*

**Abstract.** We present the eigenmodal analysis techniques enhanced towards calculations of optical and non-interacting Bose-Einstein condensate (BEC) modes formed by random potentials and localized by Anderson effect. The results are compared with the published measurements and verified additionally by the convergence criterion. In 2-D BECs captured in circular areas, the randomness shows edge localization of the high-order Tamm-modes. To avoid strong diffusive effect, which is typical for BECs trapped by speckle potentials, a 3-D-lattice potential with increased step magnitudes is proposed, and the BECs in these lattices are simulated and plotted.

## 1. Introduction

The localization concept was proposed in 1958 by P. W. Anderson who explained the absence of diffusion of waves in a disordered medium [1]. It is now recognized that this finding is a general wave phenomenon, which has been experimentally confirmed for the light [2]-[7], electromagnetic (EM) [8],[9] sound [10] and quantum waves [11]-[13] in highly disordered matter. Several approaches have been developed to model and simulate this effect. Unfortunately, the analytical results are available only for 1-dimensional (1-D) systems [14], and the higher dimensional ones are mostly studied by numerical methods.

In 1983, the evidence of Anderson's prediction was numerically found by the transfer-matrix (TM) method in [15]. The authors applied their method to the disordered electrons in 1-D system. An eigenmodal analysis of transversally-disordered electron waveguides is published in [16],[17]. Recently, scientists have spent much effort to study the localization effect of light experimentally. In addition to the experiments, Schwartz and his co-authors [3] proposed a theoretical model and solved it numerically by split step Fourier method [18]. Moreover, the localization of BECs is also a hot topic related to Anderson's original contribution. The phenomenon was realized experimentally, but not much theoretical work has been done to describe these results. In [12], the authors gave a simple equation to exponentially fit the measurements of cold atoms. In [11], the authors solved the Gross-Pitaevskii equation by sum-rules approach [19] for the frequency behavior.

Despite a vast literature on the subject, there is no efficient, fast, and general method to provide numerical study of the Anderson localization in different applications. In this paper, an approach is proposed to study this effect for photonic and matter waves. It can be generalized to all kinds of physics which are in compliance with the wave phenomena. The method arises from the analysis of eigen-mode propagation of EM field in waveguides. The method is partly similar to the 1-D analysis in [16],[17], but it uses an iterative algorithm to solve the resulting Hamiltonian matrix instead of using cascaded circuit matrices, equivalent circuit approach and TM model [20].

The paper is organized as follows. In the Section 2, the proposed method of eigenmodal analysis is briefly discussed. To demonstrate its application possibilities, the method is employed for 2-D photonic lattices in the Section 3, and for BECs in speckle potentials in the Section 4.1.

The numerical results are compared with the known measurements, and it is shown that they are in good correspondence with them. The Section 4.2 proposes a way of trapping BECs with random potential, and the Anderson localization leads to increased accuracy of trapped matter control. The Section 4.3 studies the edge states in a 2-D circular area and the influence of the potential randomness on the shapes of trapped matter.

## 2. Algorithms for the Eigenmodal Analysis

The physical systems discussed in this paper are related to the wave phenomena and described by the frequency-dependent eigen-equations of the Schrödinger type:

$$\omega v(x,y,z) = \left[-\frac{1}{2}\nabla^2 + V(x,y,z)\right] v(x,\mathrm{y},\mathrm{z}) \qquad (1)$$

where $\omega$ is the eigen frequency, $v$ is the unknown wavefunction which satisfies the normalization condition $\int_\Omega v^2 = 1$. Here, $\Omega$ is the domain of consideration. The operator $\nabla^2$ is the Laplace one, and $V(x,y,z)$ is the potential. For trapped particles, (1) gives infinite number of eigenmodes $v_i$ and its frequencies $\omega_i$. If it is necessary, the time-dependent solution is found in this case as $u(x,y,z,t) = \sum_{i=1}^{\infty} a_i v_i(x,y,z) \exp(-j\omega_i t)$ where the coefficients $a_i$ are obtained from the initial conditions at $t=0$. Here and below $j=\sqrt{-1}$ .

The Hamiltonian $H = \left[-\nabla^2/2 + V(x,y,z)\right]$ in (1) can be transformed to a square matrix by a differentiating scheme for its solution. As an example, if one considers a 1-D system and a three-step central differencing scheme and supposes that all boundary conditions are of the Dirichlet type, then eq. (1) can be written as

$$\omega\begin{bmatrix} v_1 \\ v_2 \\ v_3 \\ \vdots \\ \vdots \\ v_{N-1} \\ v_N \end{bmatrix} = \begin{bmatrix} \frac{1}{\Delta x^2}+V_1 & -\frac{1}{2\Delta x^2} & 0 & \cdots & 0 \\ -\frac{1}{2\Delta x^2} & \frac{1}{\Delta x^2}+V_2 & -\frac{1}{2\Delta x^2} & \ddots & \vdots \\ 0 & -\frac{1}{2\Delta x^2} & \frac{1}{\Delta x^2}+V_3 & \ddots & 0 \\ \vdots & \ddots & \ddots & \ddots & -\frac{1}{2\Delta x^2} \\ 0 & \cdots & 0 & -\frac{1}{2\Delta x^2} & \frac{1}{\Delta x^2}+V_N \end{bmatrix} \begin{bmatrix} v_1 \\ v_2 \\ v_3 \\ \vdots \\ \vdots \\ v_{N-1} \\ v_N \end{bmatrix}. \tag{2}$$

Here, we have used $N$-point grid in the 1-D computation domain with the spatial step $\Delta x$. The space-dependent potential $V$ is discretized as $\mathbf{V} = [V_1, V_2 \cdots\cdots V_N]^T$. The eigen frequency $\omega_n$ corresponds to an eigenvalue of the Hamiltonian matrix, and $\mathbf{v}_n = [v_1, v_2, v_3 \cdots\cdots v_N]^T$ is the corresponding eigenvector whose components represent the values of the space-dependent wavefunction. The normalization condition for the wavefunction vector is equivalent to $\sum_{i=1}^{N} v_i^2 \Delta x = 1$. If the eigenvalues of the linear system (2) are arranged as $\omega_1 < \omega_2 < \ldots < \omega_i < \ldots < \omega_N$, the eigenvector corresponding to the smallest eigenvalue $\omega_1$ is the main mode, and the eigenvector corresponding to $i^{\text{th}}$ eigenvalue $\omega_i$ is the $i^{\text{th}}$ mode.

### 2.1. Iterative Algorithms

Many algorithms are known to solve eq. (2). Due to the large size of the system, especially, in 3-D cases, only iterative algorithms are considered in this work. One option is to use the modified Rayleigth Quotient iteration [21] to compute a single eigenvalue and corresponding eigenvector, and this algorithm is shortly described below.

Modified Rayleigh Quotient iteration

*Start:*

- Choose an initial eigenvector $\mathbf{v}^0$ with $\|\mathbf{v}^0\|_2 = 1$; usually $\mathbf{v}^0$ is a random vector.

- Choose an initial eigenvalue $\omega^0$ and compute the matrix $\tilde{A}^0 = \tilde{H} - \omega^0 \tilde{I}$.
- For $n = 1, 2, 3, \cdots$ until converge:
  - Solve $\tilde{A}^{n-1}\mathbf{x}^n = \mathbf{v}^{n-1}$ for the vector $\mathbf{x}^n$.
  - Update: $\mathbf{v}^n = \mathbf{x}^n / \left\|\mathbf{x}^n\right\|_2$
  - Update: $\omega^n = \left(\mathbf{v}^n\right)^T \cdot \tilde{H} \cdot \mathbf{v}^n$
  - Update: $\tilde{A}^n = \tilde{H} - \omega^n \tilde{I}$

*End.*

Here, $\tilde{H}$ is the Hamiltonian matrix in (2), $\tilde{I}$ is the identity matrix, and $\| \ \|_2$ is the second norm of vectors. Note, we have modified the standard Rayleigh Quotient iteration in [21], so that the obtained results are largely dependent on the initial guess of $\omega^0$, i.e. the algorithm used in our work converges to an eigenpair, and the obtained eigenvalue is close to the initial guess $\omega^0$. The final obtained eigenvector $\mathbf{v} = \left[v_1, v_2, \cdots, v_N\right]^T$ satisfies $\left\|\mathbf{v}\right\|_2 = 1$, therefore it needs to be divided by $\sqrt{\Delta x}$ to allow the normalization condition $\sum_{i=1}^{N} v_i^2 \Delta x = 1$. The main advantage is that this algorithm is always convergent at a cubic rate since the matrix $\tilde{H}$ is real and symmetric. However, a disadvantage is that the initial guess $\omega^0$ should be close enough to the desired eigenvalue.

Another option is to use the Krylov-subspace method to compute the eigenvalues and corresponding eigenvectors. For example, one can use the Matlab built-in command '*eigs*' to compute modes. This command employs the implicitly restarted Arnoldi iteration [21], and it gives the possibilities to compute several smallest or several largest modes. It is therefore not necessary to choose the initial guess carefully as in the Rayleigh Quotient case. Unfortunately, this method does not always work well according to our experience. In some cases, it diverges due to the ill-condition of the matrix caused by randomness of the spatially-dependent potential.

In general, there is no single algorithm that fits best for all kinds of applications. The authors of this paper conclude that the Rayleigh Quotient iteration better fits the photonic case (Section 3), and the Arnoldi iteration is more suitable for the BEC cases (Section 4). On the other hand, if one tries to solve the photonic case with Krylov-subspace method, it is very likely that the solution diverges.

### 2.2. Precision of the Numerical Solutions

The accuracy of numerical solutions depends on the grid. Roughly speaking, the numerical solutions become more precise as the number of grid points approaches infinity. Due to the limitation of computer memory, it is only possible to use the grids of limited size, and an approach to evaluate the accuracy of numerical solutions needs to be obtained.

We started by checking the case when the potential is harmonic, i.e. $V = V_0 \sum_{k=1}^{K} \sin^2 \omega_s x_k$ , where $x_k$ is the variable of $k^{\text{th}}$ axis in a $K$-dimensional system, $\pi / \omega_s$ is the periodicity, and $V_0$ is the amplitude of the potential. A direct observation of the accuracy is to compare the numerical solutions with the analytical ones. Unfortunately, analytical solutions are not available in higher dimensions (2-D and 3-D) due to mathematical difficulties. Therefore, an implicit method is used to analyze the numerical accuracy in this paper.

The quantity $\epsilon_r = \int_\Omega \left| \Psi_{\text{Num}}^N - \Psi_{\text{A}} \right|^2 / \int_\Omega \left| \Psi_{\text{A}} \right|^2$ will indicate the error (in percentage) in the numerical solution regarding to the analytical one within the considered domain $\Omega$. Here, $\Psi_{\text{Num}}^N$ indicates the numerical solution obtained from $N$-point grid (total points are $N$ in 1-D, $N^2$ in 2-D, and $N^3$ in 3-D), and $\Psi_{\text{A}}$ is the analytical solution. Consider the numerical solution $\Psi_{\text{Num}}^N$ and

$\Psi_{\text{Num}}^{N+1}$ derived from $N$-point and $N$+1-point grid. If these three vectors $\Psi_{\text{Num}}^{N}$, $\Psi_{\text{Num}}^{N+1}$, and $\Psi_{\text{A}}$ are expressed in a multi-dimensional space, the triangle law implies

$$\left|\Psi_{\text{Num}}^{N+1}-\Psi_{\text{A}}\right|+\left|\Psi_{\text{Num}}^{N+1}-\Psi_{\text{Num}}^{N}\right|\geq\left|\Psi_{\text{Num}}^{N}-\Psi_{\text{A}}\right|. \tag{3}$$

Taking a square of both sides of eq. (3), one obtains,

$$\left|\Psi_{\text{Num}}^{N+1}-\Psi_{\text{A}}\right|^{2}+2\left|\Psi_{\text{Num}}^{N+1}-\Psi_{\text{A}}\right|\left|\Psi_{\text{Num}}^{N+1}-\Psi_{\text{Num}}^{N}\right|+\left|\Psi_{\text{Num}}^{N+1}-\Psi_{\text{Num}}^{N}\right|^{2}\geq\left|\Psi_{\text{Num}}^{N}-\Psi_{\text{A}}\right|^{2}. \tag{4}$$

Assuming that $N$+1 is large enough so that $\Psi_{\text{Num}}^{N+1}\to\Psi_{\text{A}}$, the first and second term on the left hand side of (4) is close to zero and the inequality is reduced to

$$\left|\Psi_{\text{Num}}^{N+1}-\Psi_{\text{Num}}^{N}\right|^{2}\geq\left|\Psi_{\text{Num}}^{N}-\Psi_{\text{A}}\right|^{2}. \tag{5}$$

By taking integral over the interested domain $\Omega$ on both sides of (5) and then dividing by $\int_{\Omega}\left|\Psi_{\text{A}}\right|^{2}$, one can rewrite this inequality as

$$\frac{\int_{\Omega}\left|\Psi_{Num}^{N+1}-\Psi_{\text{Num}}^{N}\right|^{2}}{\int_{\Omega}\left|\Psi_{\text{A}}\right|^{2}}\geq\frac{\int_{\Omega}\left|\Psi_{\text{Num}}^{N}-\Psi_{\text{A}}\right|^{2}}{\int_{\Omega}\left|\Psi_{\text{A}}\right|^{2}}=\epsilon_{r}. \tag{6}$$

Using the normalization condition for the analytical solution $\int_{\Omega}\left|\Psi_{\text{A}}\right|^{2}=1$, one finally obtains

$$\int_{\Omega}\left|\Psi_{\text{Num}}^{N+1}-\Psi_{\text{Num}}^{N}\right|^{2}\geq\epsilon_{r}. \tag{7}$$

Even if the analytical solution is unavailable, it is still possible to know the relative error $\epsilon_r$ by checking the quantity $\int_{\Omega}\left|\Psi_{\text{Num}}^{N+1}-\Psi_{\text{Num}}^{N}\right|^{2}$ according to (7) because $\epsilon_r$ is bounded by it. It is assumed that the quantity $\int_{\Omega}\left|\Psi_{\text{Num}}^{N+1}-\Psi_{\text{Num}}^{N}\right|^{2}$ shows a decaying behavior because both $\Psi_{\text{Num}}^{\text{N}}$ and

$\Psi_{\text{Num}}^{N+1}$ are close to the analytical solution, and the difference between them becomes less when the grid becomes finer.

The numerical experiments are performed for 1-D, 2-D, and 3-D cases. The magnitude of the periodic potential is chosen to be relatively large $V_0 = 1000$, and ten periods of potential is included in the interested domain. We simulate for *N* up to 500 in 1-D and 2-D, and *N* up to 88 only in 3-D due to the administrative limitation on the used supercomputer memory resources. The results are illustrated in Fig. 1.

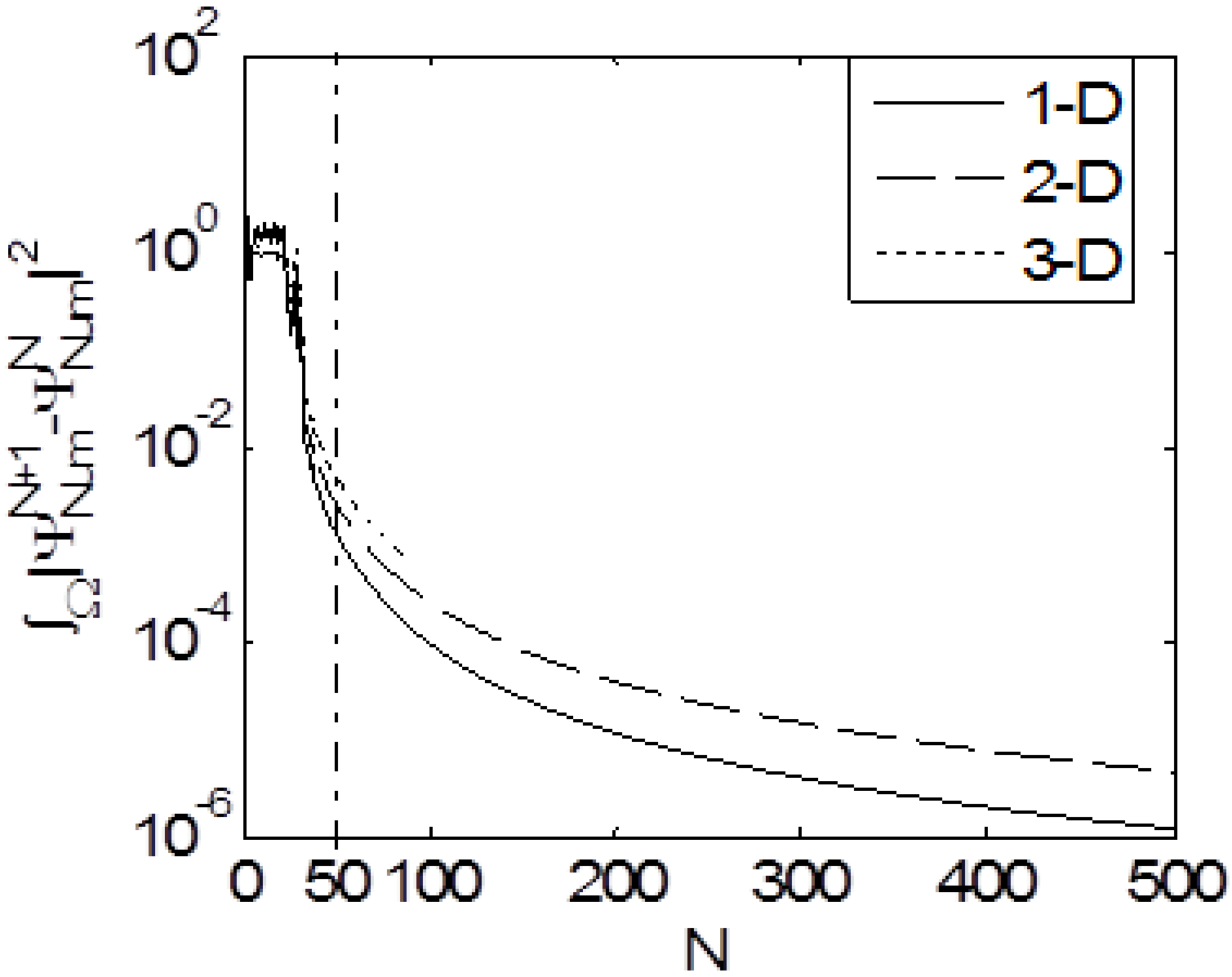

Fig. 1. The quantity $\int_\Omega \left| \Psi_{\text{Num}}^{N+1} - \Psi_{\text{Num}}^{N} \right|^2$ as a function of grid points. Simulations are performed for 1-D, 2-D and 3-D cases, and ten periods of potential is included along each axis of the interested domain. Solid line: 1-D case; Dashed line: 2-D case; Dotted line: 3-D case.

It is obvious that the higher dimensional simulations introduce more error to the numerical solutions due to some instability of algorithm which typical for many iterative methods [15]. We specially draw a vertical dash-dot line at *N*=50 in Fig. 1 to show the safety choice of the number of grid points. When *N* is greater than 50, the quantity $\int_\Omega \left| \Psi_{\text{Num}}^{N+1} - \Psi_{\text{Num}}^{N} \right|^2$ is below $10^{-2}$ even for 3-D cases. This means that the relative error $\epsilon_r$ is less than 1%. Since we include ten periods of potential in all cases, $N \geq 50$ is equivalent to at least five points for each period of potential.

It is concluded that the number of grid points $N$ should be chosen according to the space-dependent potential in (1). We are going to study the localization effect for different application cases in the next two sections. The precision of the numerical solutions is very difficult to derive due to the randomness of potentials. Therefore, for each case in the Sections 3 and 4, an average value of the quantity $\int_\Omega \left|\Psi_{\mathrm{Num}}^{N+1} - \Psi_{\mathrm{Num}}^{N}\right|^2$ is calculated through 100 samples, so that general idea of relative error is applicable.

## 3. Application to 2-D Photonic Lattices

We first use the eigenmodal analysis to simulate the 2-D photonic lattices. The authors in [3] did an experiment to study the Anderson localization of 514-nm laser beam. They used a hexagonal photonic lattice with a periodicity of 11.2 μm and a refractive-index $n_0 = 2.34$. The authors employed an optical induction technique to generate random fluctuation upon refractive-index of the lattice, so that this refractive-index and its fluctuation of the lattice can be regarded as 'random periodic potential'. The light beam propagates in the $z-$direction along which the lattice is uniform, and the authors claim that the system is described by a Schrödinger-like equation:

$$j\frac{\partial A}{\partial z} + \frac{1}{2k}\left(\frac{\partial^2 A}{\partial x^2} + \frac{\partial^2 A}{\partial y^2}\right) + \frac{k}{n_0}\Delta n(x, y)A = 0. \tag{8}$$

Here, $A(x, y, z)$ is related to the optical intensity $I = |A|^2$, $\Delta n(x, y)$ is the random fluctuation upon the lattice, $k = 2\pi f n_0 / c$ with $f$ as the frequency, and $c$ is the speed of light in vacuum. The authors in [3] used a generalized method to solve eq. (8), namely, they treat the axis $z$ as time and simulate (8) with the split-step Fourier method.

The eigenmodal analysis assumes that $A = \Phi(x, y)\exp\left(j\left(k_z z - 2\pi ft\right)\right)$ is the solution form for (8) with $k_z$ as the propagation constant in $z-$direction. Plugging this formula into eq. (8) and using $k = 2\pi fn_0 / c$, one obtains

$$\frac{1}{2}\left(\frac{\partial^2\Phi}{\partial x^2} + \frac{\partial^2\Phi}{\partial y^2}\right) + \frac{4\pi^2}{n_0\lambda^2}\Delta n(x, y)\Phi = \frac{2\pi k_z}{\lambda}\Phi. \tag{9}$$

Here, we have used $\lambda = c/n_0 f$ as the wavelength of the laser light. Note the eq. (9) is now $z-$independent. Our method is more efficient than the split-step Fourier one as it was found from our simulations because it avoids solving the $x-y$ planes along the light propagation direction.

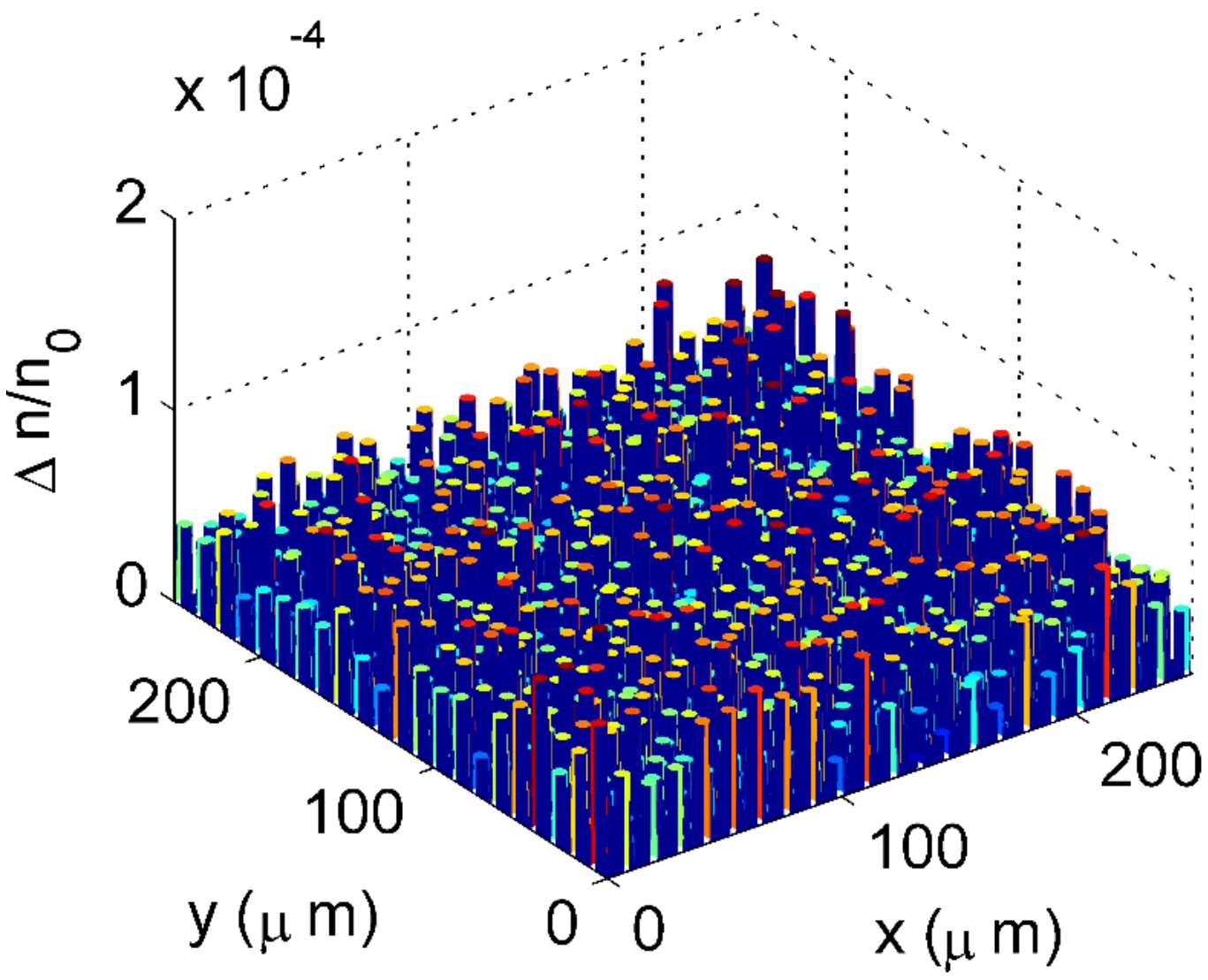

Fig. 2. A typical representation of $\Delta n / n_0$ in the x-y plane in a 250×250 μm area; Periodicity of the potential is 11.2 μm; Its magnitude is random and its disorder level $d$ is 15%.

The ratio between the random fluctuations and refractive index is very small ($|\Delta n| / n_0 < 2\times10^{-4}$) according to [3]. Therefore, random periodic fluctuation is modeled by:

$$\frac{\Delta n(x, y)}{n_0} = \begin{cases} 0 & \text{if } (x,y) \text{ is outside the lattice,} \\ 10^{-4} r & \text{if } (x,y) \text{ is inside the lattice.} \end{cases} \tag{10}$$

Here $r$ is a random number in the interval $[0,2]$ and it is of the normal distribution. Its statistical characteristics are the mean value $\bar{\mu}=1$ and standard deviation $\sigma=d^{1/2}$ , where $0<d<1$ is the level of disorder (normalized in percentage) regarding the mean value $\bar{\mu}$. It should be mentioned that (10) is just an approach to approximate the random fluctuation of optical lattice due to the optical induction technique used in [3]. It matches the condition $|\Delta n|/n_0<2\times10^{-4}$ which the authors of [3] used in their experiments. A typical plot of spatially dependent $\Delta n/n_0$ is shown in Fig. 2.

The computation is performed in a 2-D area $[0,250]\times[0,250]$ μm with the Dirichlet boundary condition on all boundaries. We perform simulations on an $N$=1400 2-D grid. The average relative error in this case is 3.01%. To visualize the localization effect, we check the smallest eigenvalue of (9) and its corresponding eigenvector of the Hamiltonian matrix. Localization of this highest propagating mode is the sign that all lower modes have been already localized in these conditions [17]. For illustration purpose, the optical intensity distributions at the disorder levels 0%, 15%, and 45% are plotted in the $x-y$ plane and are compared with the measured ones from [3] where the optical field is shown for localized beam composed of many propagating modes. Then, our numerical results for the highest propagating mode (Fig. 3) visually are in good correspondence with the mentioned measurements.

Since the random numbers are generated by computer, it is important to measure the averages of samples over many realizations. We simulate eleven different levels of disorder, and we calculate the average of the effective width $w_{\text{opt}}=\left\{\left[\int I(x,y)^2\,dxdy\right]/\left[\int I(x,y)\,dxdy\right]^2\right\}^{-1/2}$ from 100 realizations.

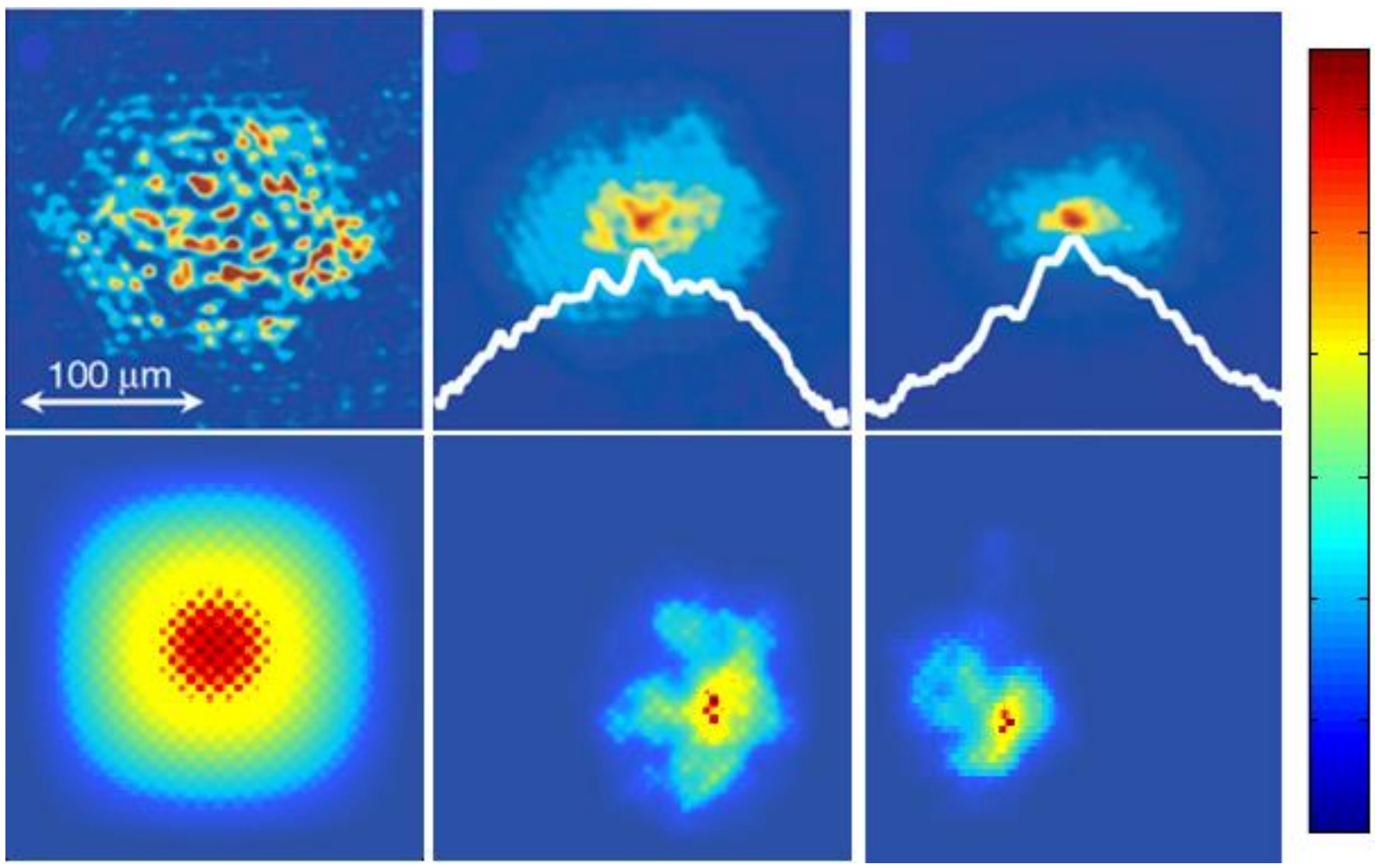

Fig. 3. Intensity distribution in the $x-y$ plane in 250×250 μm area regarding different disorder levels $d$. Computation is performed on the 1400×1400 grid. The upper row: beam measurements adapted from [3]; the upper row from the left to right are for 0%, 15%, and 45% disorder level $d$. Lower part: Numerical results by our eigenmodal analysis approach given for the highest propagating mode; the lower row from the left to right are for 0%, 15%, and 45% disorder level.

Fig. 4 shows the averaged effective width $w_{\text{opt}}$ as a function of disorder level given for the highest propagating mode of (9). The beam measurements from [3] (stars) are used for comparisons. Our numerical results (square marks in Fig. 4) show that the modal effective width $w_{\text{opt}}$ decreases exponentially as the level of disorder increases, and it shows correspondence to beam measurements (star marks in Fig. 4) confirming that the highest propagating mode can define the beam width. The simulated by split step Fourier method beam width from [3] (circles in Fig. 4) is also included in this figure.

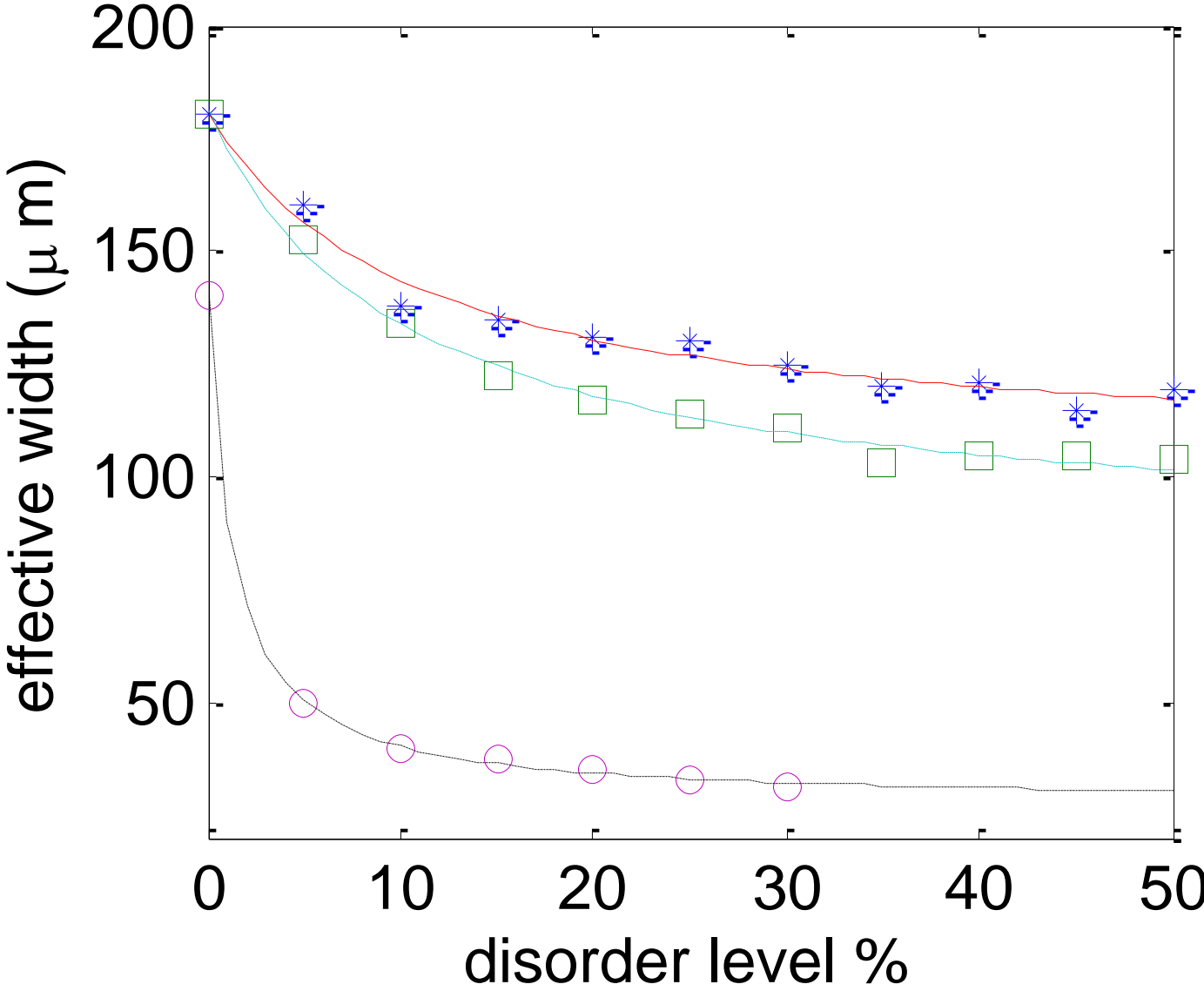

Fig. 4. Ensemble averaged effective width as a function of disorder level. Square marks: modal effective width calculated by eigenmodal analysis taken from the average of 100 realizations, computation is performed in 250×250 μm area with the 1400×1400 grid; Star marks: beam measurements from [3] taken from the average of 100 measurements; Circles: beam effective width $w_{\text{opt}}$ calculated by the split step Fourier method from [3] taken from the average of 100 realizations; Solid curve: least-square curve fitting for the star marks; Dashed curve: least-square curve fitting for the square marks; Dash-dotted curve: least-square curve fitting for the circles.

The application of eigenmodal analysis to the photonic case is demonstrated in this Section. In addition, the main advantage of this approach in photonic case is that it avoids solving numbers of $x-y$ planes one after another along the propagation direction. This simulation advantage will become more obvious when light propagates for longer distances.

## 4. Application to Non-interacting BECs

We now study the localization effect in non-interacting BECs [22],[23],[24]. The time-independent Gross-Pitaevskii equation is used to model this condensate in the external potential $V_{\text{ext}}(x,y,z)$:

$$\mu\Psi(x,y,z)=\left[\frac{\hbar^2}{2m}\nabla^2+V_{\text{ext}}(x,y,z)\right]\Psi(x,y,z). \tag{11}$$

Here, $\mu$ is the chemical potential as the eigenvalue, $\Psi(x,y,z)$ is the wave function, $\hbar$ is the normalized Planck constant, and $m$ is the mass of atom (Rb87 in our case). The atom-atom interaction is omitted due to the non-interacting assumption.

Using the eigenmodal analysis method and calculating the smallest eigenvalue and its corresponding eigenvector (the main mode) of (11), one is able to study the behavior of BECs regarding the lowest chemical potential $\mu$.

### 4.1. 3-D Non-interacting BECs in Speckle Potential

The speckle pattern is produced by the mutual interference of a set of optical wave fronts [25]. Ever since the invention of the laser, the speckle pattern finds a variety of applications. For instance, the authors in [12] have employed an optical speckle field to the BECs and found the Anderson localization effect. Optical speckle fields are also used to trap ultra-cold Fermi gas in [26].

In our particular case, the computation domain is $[-50,50]\times[-50,50]\times[-50,50]\,\mu\text{m}$, and the simulation is performed on the *N*=50 3-D grid. The average relative error (7) in this case is 3.72%. The boundary conditions are of the Dirichlet type on all boundaries. A 3-D speckle type random potential is first generated according to the method in [27], as illustrated in Fig. 5. It has the same characteristics as the speckle potential in Fig. 1-(a) from [26].

We are interested to know the disorder of speckle potential. Unlike the case in Section 3, the disorder level cannot be normalized regarding the mean value of the generated speckle potential, since the speckle generation algorithm has no possibility to control it. Therefore $V_{\text{STD}}$ is defined as the standard deviation of the speckle potential, and it indicates the magnitude of the disorder (not in normalized form). When $V_{\mathrm{STD}} / h = 0$ ( $h$ is the Planck constant), it is seen that the trapped BECs form a smooth sphere in Fig. 6-(a); As the standard deviation increases to $V_{\mathrm{STD}} / h = 50$, the shape of the BECs starts to be distorted due to, presumably, the Anderson effect (Fig. 6-(b)); When the disorder of the speckle potential continues increasing, the trapped BECs start to split in Fig. 6-(c). This conclusion is in conformity with the reported measurements [11], which are illustrated in the insets of corresponding figures.

In order to quantify the localization effect in this case, the effective width of trapped BECs is defined as $w_{\mathrm{BEC}} = \left\{ \left[ \int \Psi^2 dxdydz \right] / \left[ \int \Psi dxdydz \right]^2 \right\}^{-1/2}$. For statistical reasons, we take 11 different $V_{\mathrm{STD}} / h$ from 0 to 2000 and simulate 100 realizations for each. The effective width averaged from the 100 realizations is showing exponential decreasing regarding $V_{\mathrm{STD}}$ (Fig. 7). This conclusion is in correspondence with the published 1-D measurements presented in Fig. 3 of [12]. Like the photonic case in Section 3, the Anderson effect possibly results in the localization of BECs, and this phenomena is successfully modeled by our approach.

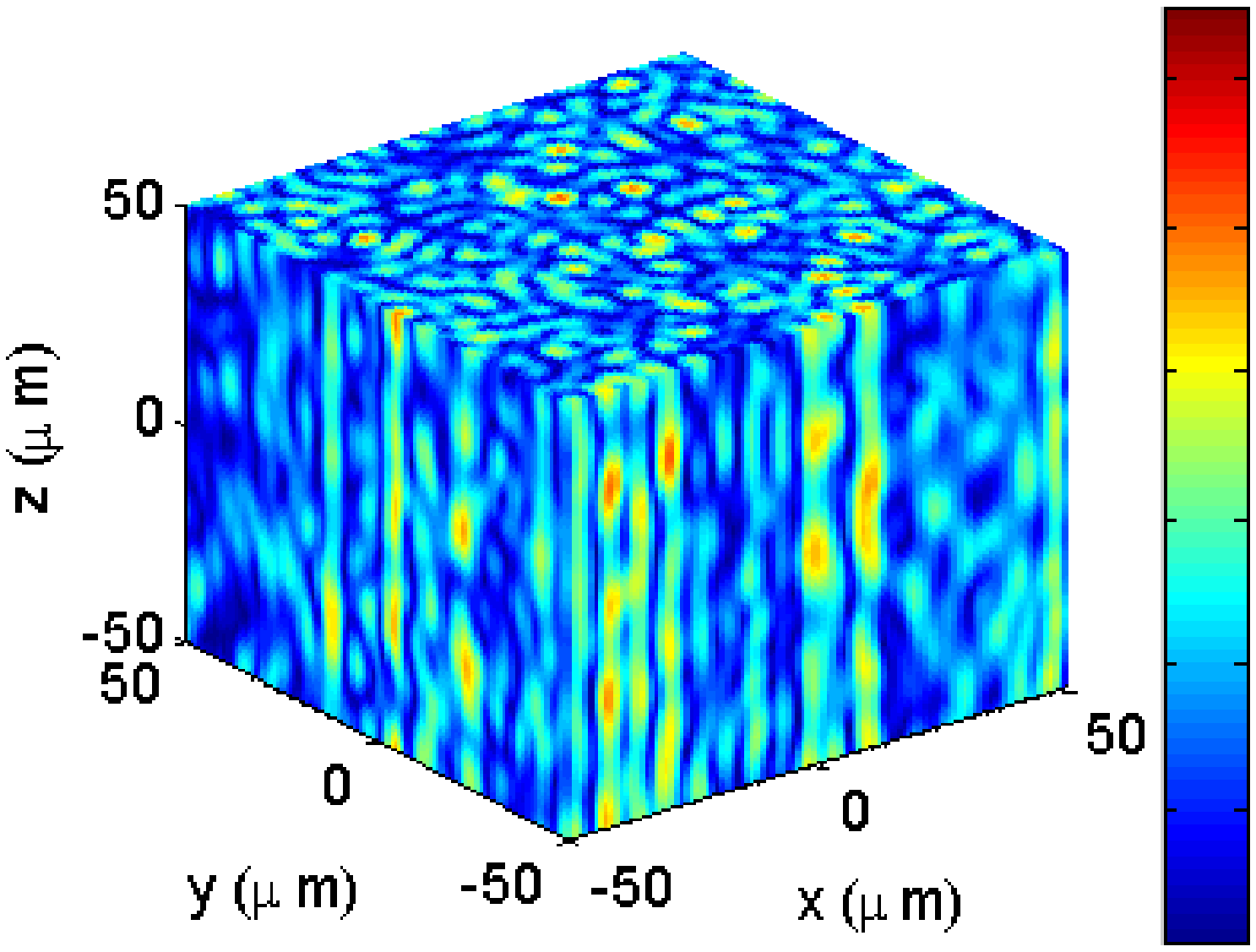

Fig. 5. A typical example of slice plot of speckle pattern random potential at x=-50, y=-50, and z=50 μm planes. The potential is generated by computer according to the method in [27].

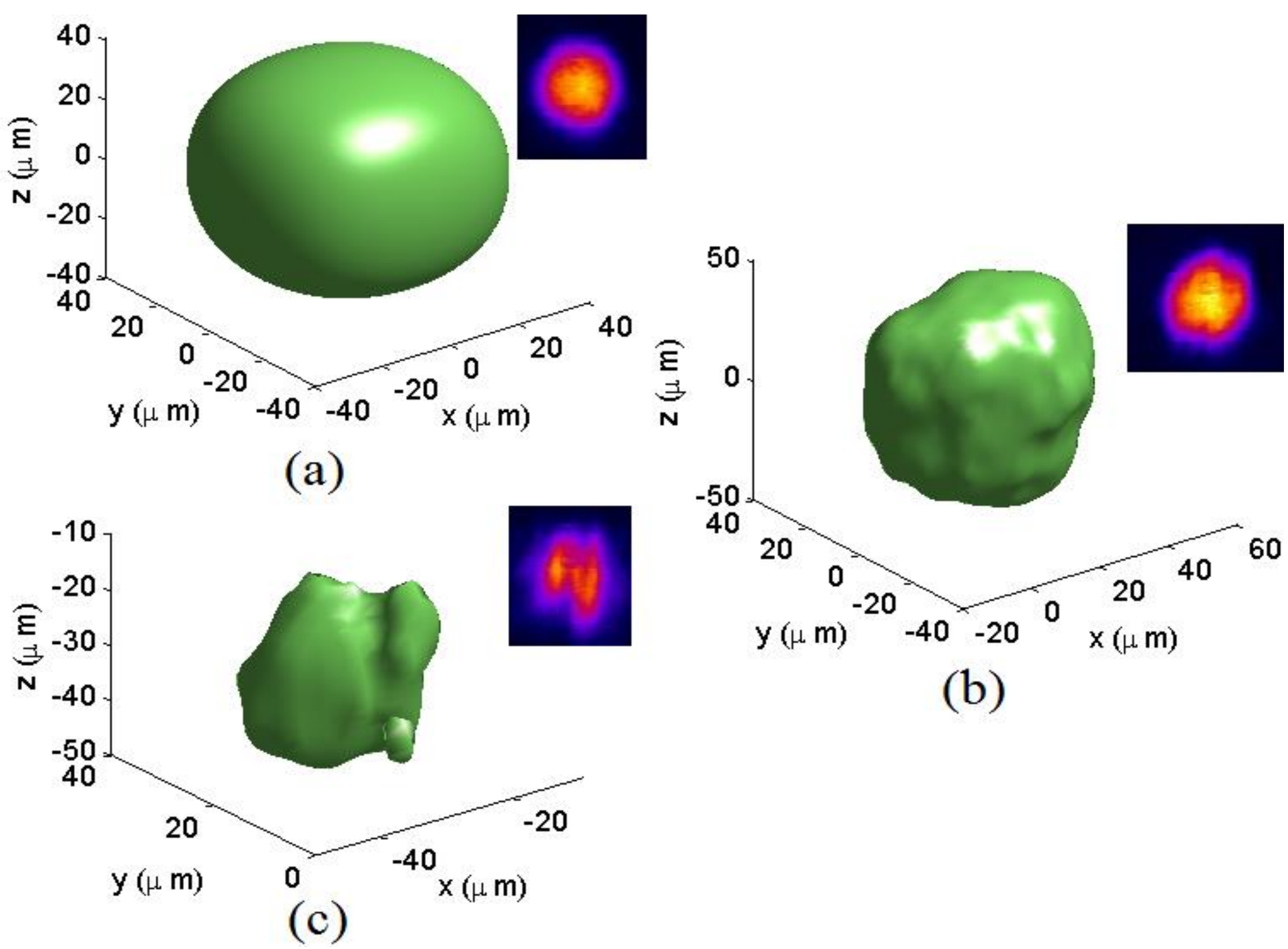

Fig. 6. 3-D surface plot of the density profile of BECs with respect to (a) standard deviation of speckle potential $V_{\mathrm{STD}} = 0$ ; (b) $V_{\mathrm{STD}} = 50$ ; (c); $V_{\mathrm{STD}} = 300$ . The insets are the 2-D measured absorption images adopted from [11] (APS permission #3884261326413). These simulations are performed in a 100×100×100 μm area with the 50×50×50 grid.

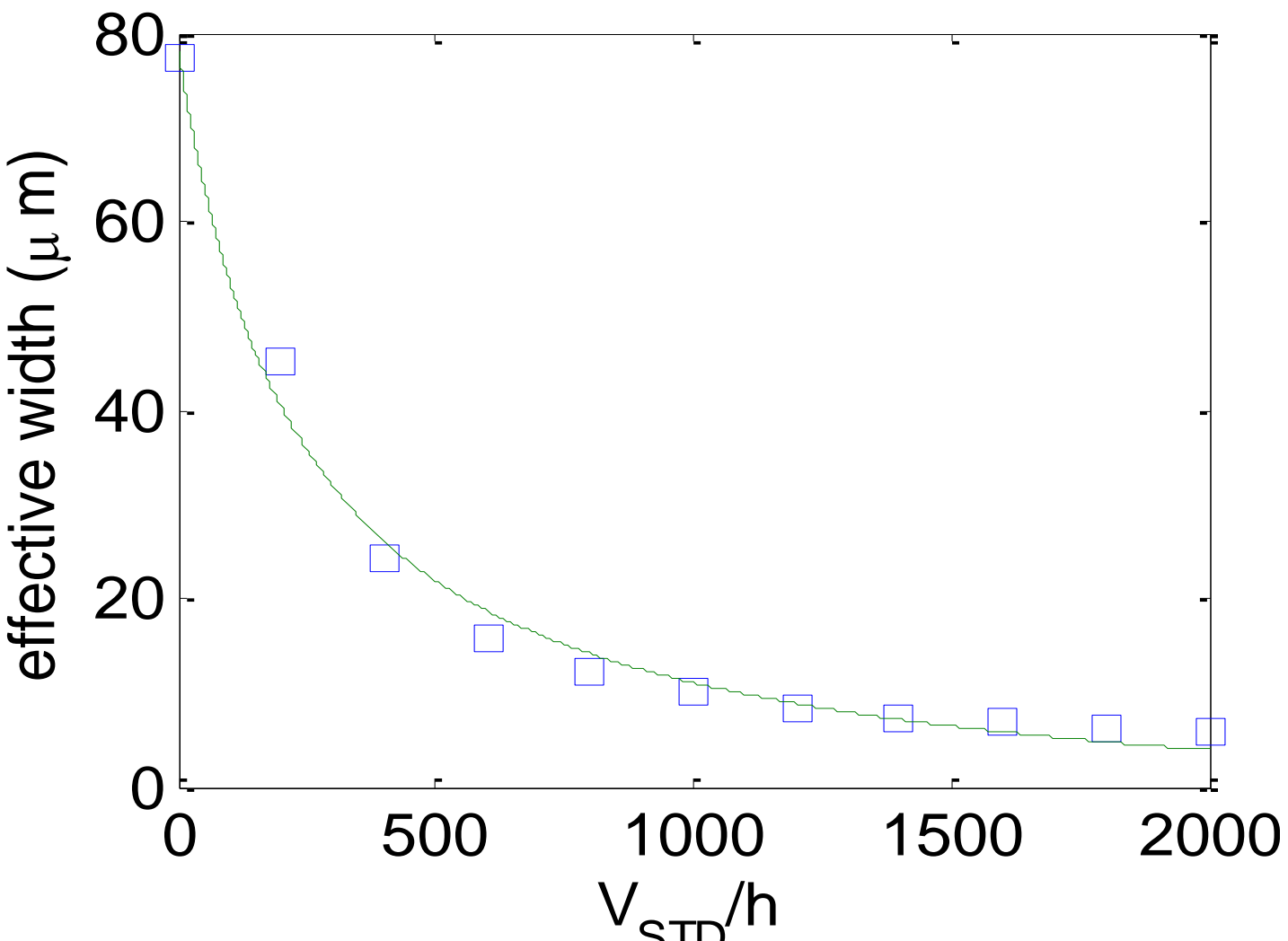

Fig. 7. Simulated effective width $w_{\text{BEC}}$ of BECs versus the standard deviation $V_{\text{STD}}$ of speckle potential. Computation domain is an area of 100×100×100 μm, and the grid is 50×50×50. Squares: calculated $w_{\text{BEC}}$ averaged by 100 realizations; solid line: least square curve fitting for square marks.

### 4.2. Non-interacting BECs Trapped in 3-D Lattice Potential

The trapping potentials in [11],[12],[26] were generated using the diffusive plates. Due the smoothness of this random potential distribution, only a part of atoms of BECs is localized while others are spreading away with the time [13]. To decrease this diffusive component of the trapped BECs, a random 3-D lattice potential magnitude [28] can be used (Fig. 8). It is periodic in the $x-$, $y-$, and $z-$ directions (approximately 20 μm periodicity), and, in general, its period is not connected to the wavelength of laser light as in [28]. According to the de Broglie theory, the matter wavelength of BECs is calculated as $\lambda = \sqrt{2\pi\hbar^2 / mK_b T}$, where $m$ is the mass of Rb87 atom, $K_b$ is the Boltzmann constant, and $T$ is the temperature in nano-Kelvin (in our case $T = 7\times10^{-9}$ K ). The wavelength of BECs is approximately several tens of micrometers. Therefore the periodicity of the lattice potential is comparable to the BEC's wavelength, and the localization effect occurs. The magnitude of potential is random, and it can be realized using holography or photomasks. Similar to the photonic lattice and the photonic speckle potentials, the disorder level of this 3-D lattice is related to the standard deviation of randomness.

Simulations of eq. (11) is performed in a 3-D area $[-50,50]\times[-50,50]\times[-50,50]\,\mu\text{m}$ with $N$=50 grid. The average relative error (7) of our calculations in this case is 5.65%. The numerical results show that the size of BECs decreases monotonically as the disorder level increases (Fig. 9). It is also shown that the average effective width $w_{\text{BEC}}$ decreases exponentially with respect to the disorder level (Fig. 10). This effective width of BECs is calculated for 11 levels of disorder from 0% to 50%, and the results for each level are taken from the average of 100 realizations. Only four of them are shown in Fig. 9 as the examples.

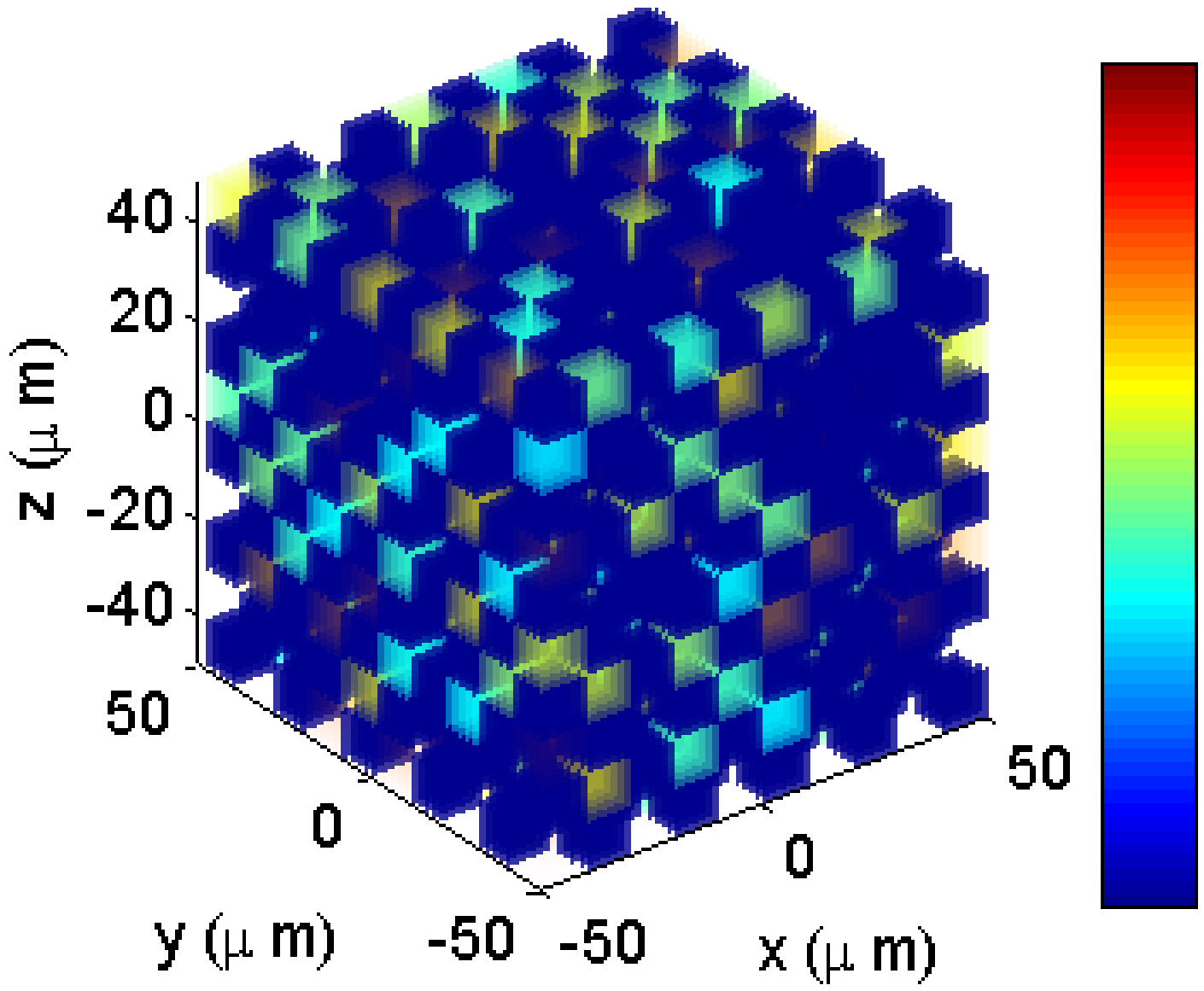

Fig. 8. A typical presentation of 3-D lattice potential in a 100×100×100 μm area. It is a 20 μm period lattice, and the disorder level is 15%. The color indicates the strength of the potential.

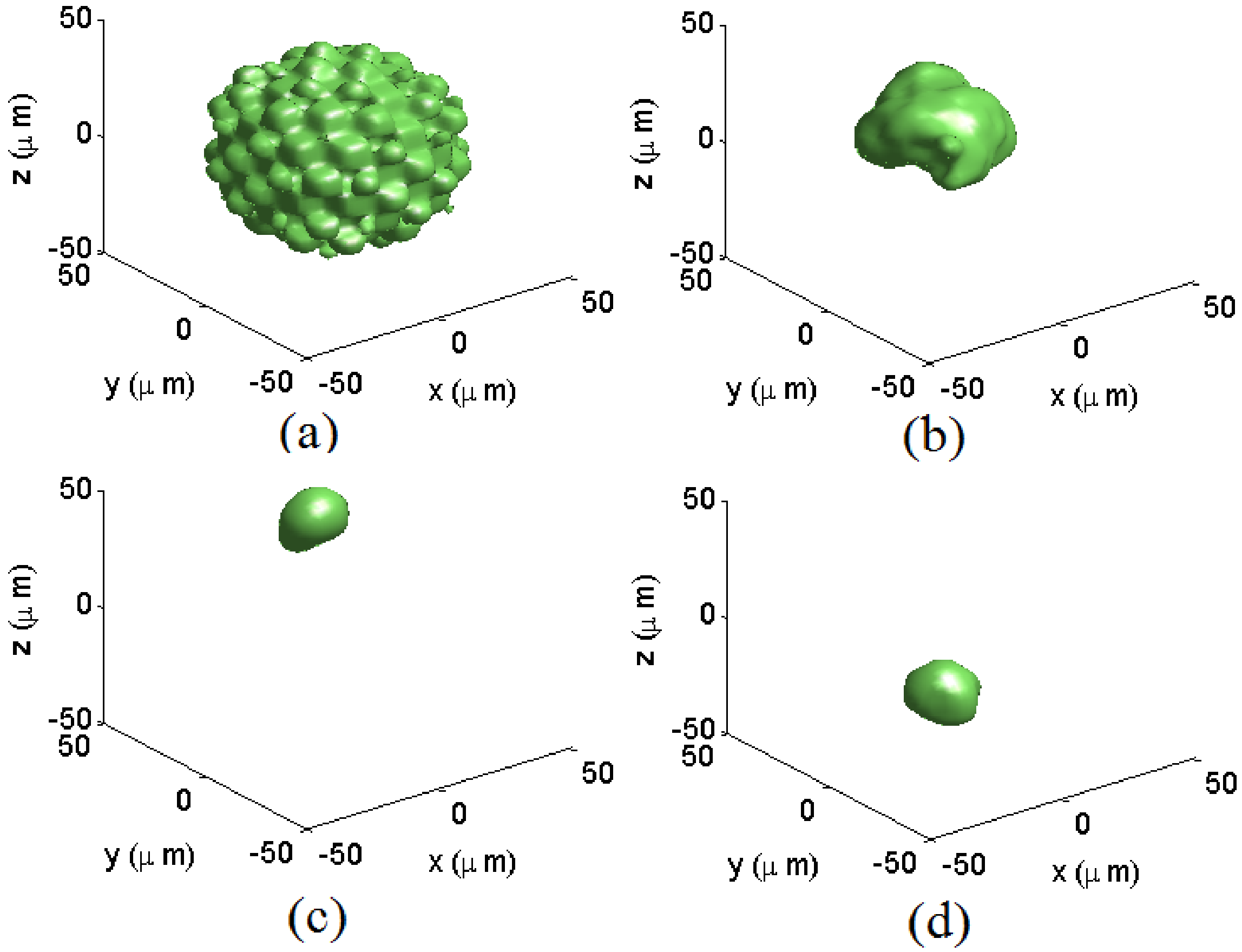

Fig. 9. Surface plots of BECs trapped by disordered lattice potential (Fig. 8) regarding different disorder level. Computation domain is an area of 100×100×100 μm and the grid is 50×50×50. (a) no disorder; (b) 10% disorder; (c) 30% disorder; (d) 40% disorder.

By controlling the disorder level, it is possible to reduce the size of the trapped BECs down to a certain limit. In our case, the effective width $w_{\mathrm{BEC}}$ stops decreasing when reaching 10 μm, and this is exactly a half period of the lattice potential. Another limit is with the known relationship of the BEC wavelength and the period of the lattice. This means that the size of BECs can be controlled by the lattice in a certain limit if enough disorder level is provided. This is an advantage of using the lattice potential comparing to the traditional method of trapping BECs (e.g. using harmonic potential).

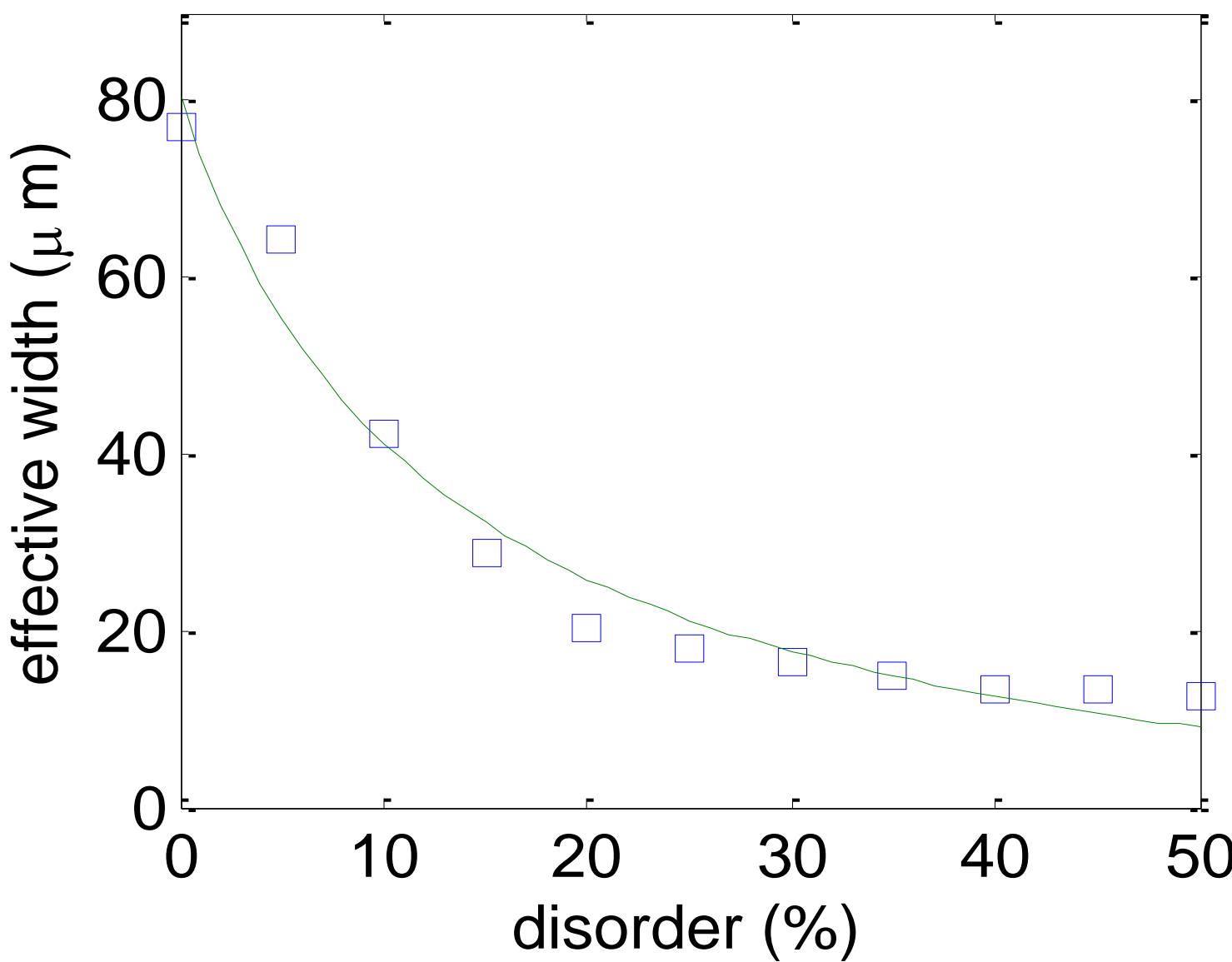

Fig. 10. Simulated effective width of BECs versus disorder level of lattice potential. Computation domain is an area of 100×100×100 μm, and the grid is 50×50×50. Squares: simulated effective width averaged from 100 realizations; Solid line: least square curve fitting of square marks.

### 4.3. Higher-order Modes of Non-interacting BECs in 2-D Lattice Potential

If one arranges the eigenvalues of eq. (11) as $\mu_1 < \mu_2 < \ldots \mu_i < \ldots$, this paper has so far focused on the smallest eigenvalue $\mu_1$ and its corresponding eigenvector. We now stay with eq. (11) for the higher-order mode calculations, namely, $\mu_i$ such that $i > 1$ and for corresponding eigenvectors. For this purpose, the BECs trapped by lattice potentials in a 200×200-μm 2-D area is simulated, and a 2-D grid of $N$=400 is used for this purpose. The average relative error of 9.03% in our calculations is registered. The lattice potential of the spatial period around 10 μm exists only in a circular area, which is centered at the original point with the radius of 90 μm. The hard-wall boundary (infinitely large potential) is applied in the surrounding margin area (Fig. 11).

Initially, consider the non-random potential. In linear optics, Tamm-like surface waves were suggested in periodic media [29]. Due to the periodicity of the used lattice potential, it is not surprising that the ring-structured BECs are found close to the boundary by simulations. In our particular case, not only single rings (the first row in Fig. 12), but also double rings (2nd row in Fig. 12) are found when the lattice potential is in absence of disorder. We notice that these rings in our simulations are periodic in angular direction, and the angular spatial frequency is proportional to the chemical potential $\mu$, e.g. for the single rings in the first row of Fig. 12, the chemical potential increases from (a) to (c), and it is seen that the angular spatial frequency also increases. This is because the chemical potential in eq. (11) corresponds to the eigen frequency in eq. (1), and the latter one determines the frequency in spatial domain.

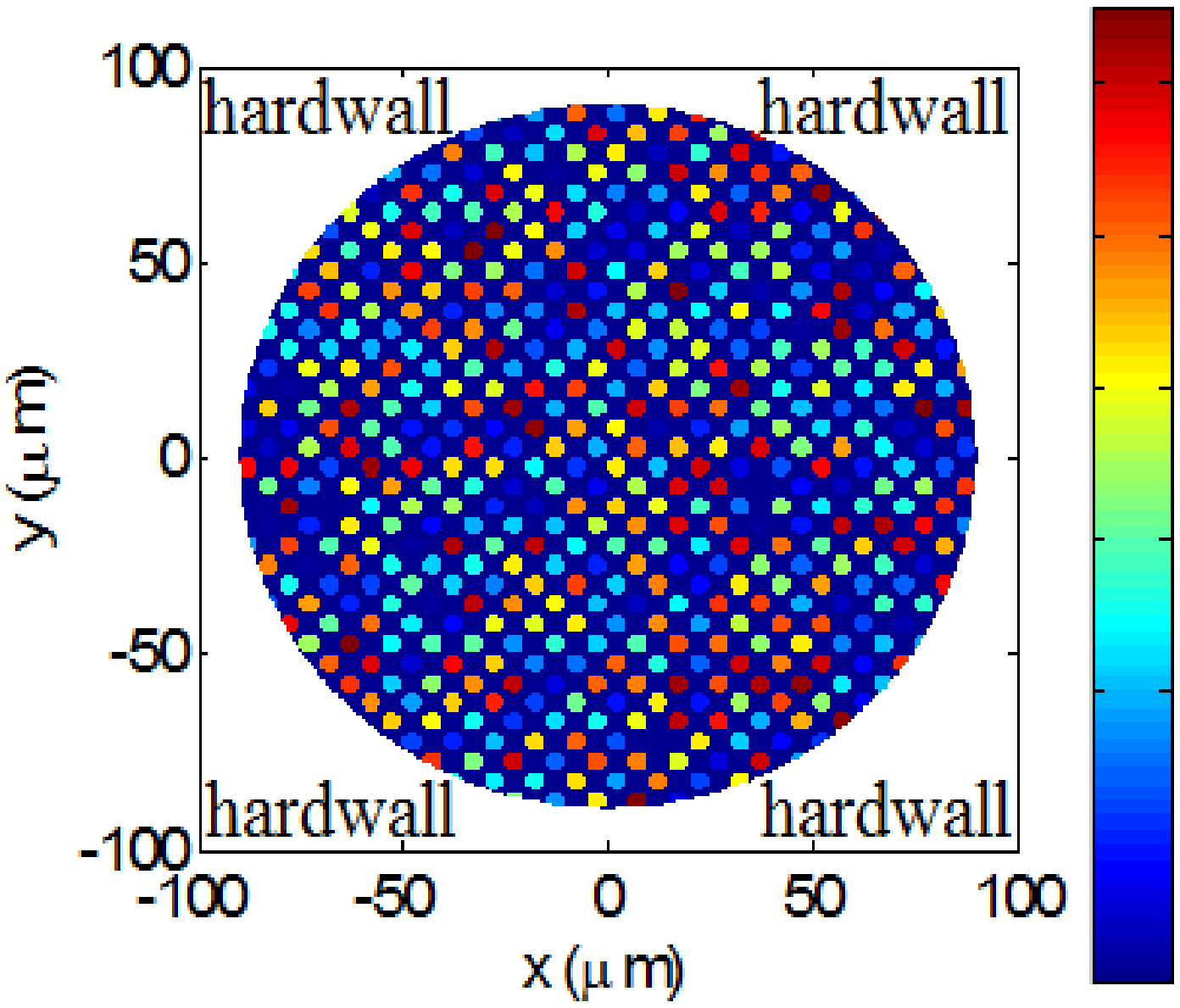

Fig. 11. 2-D random lattice potential in a circle area inside a 200×200 μm domain with 400×400 grid. The period of the lattice potential is 10 μm. The white margins around the corners are hard-wall boundary conditions.

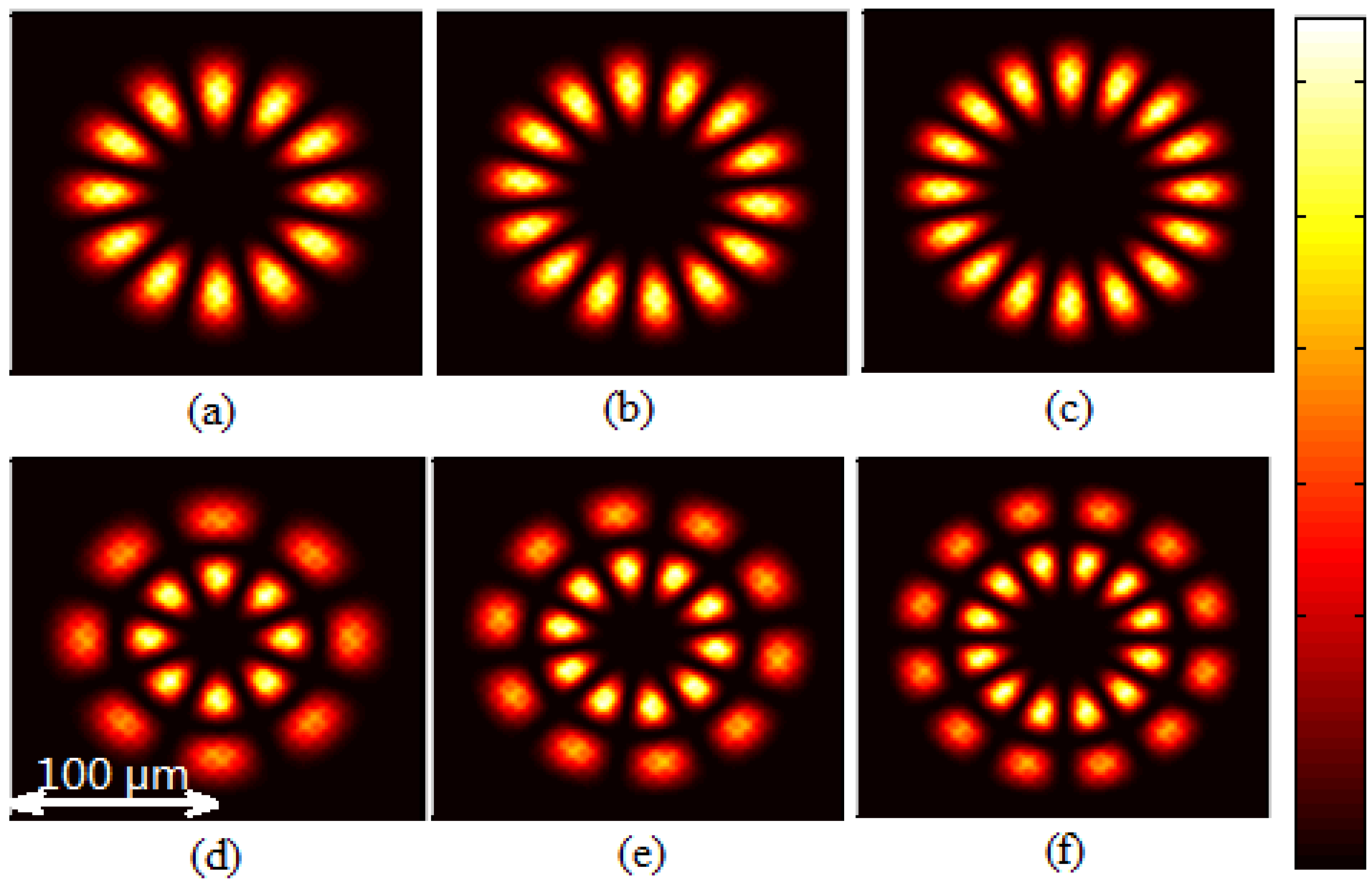

Fig. 12. Ring-structured BECs in the lattice potential (Fig. 11) with respect to chemical potential $\mu$ of different values. Computation domain is an area of 200×200 μm with the grid is 400×400. The lattice potential is without disorder for these pictures. (a) $\mu / h = 224$ ; (b) $\mu / h = 245$ ; (c) $\mu / h = 266$ ; (d) $\mu / h = 225$ ; (e) $\mu / h = 247$ ; (f) $\mu / h = 270$ .

We now focus on the chemical potential $\mu_{24}$ (the 24th eigenvalue of eq. (11)) and its corresponding eigenvector. This mode has a double ring structure when there is no disorder, and BECs are evenly divided on the inner ring and the outer ring (Fig. 13-(a)). The mean amplitude of the lattice potential is set to $\langle V_{ext} / h \rangle = 500$ , and we check 120 realizations for every disorder level at 1% (Fig. 13 (b-e), 5% (Fig. 13(f-i)), and 30% (Fig. 13 (j-m)).

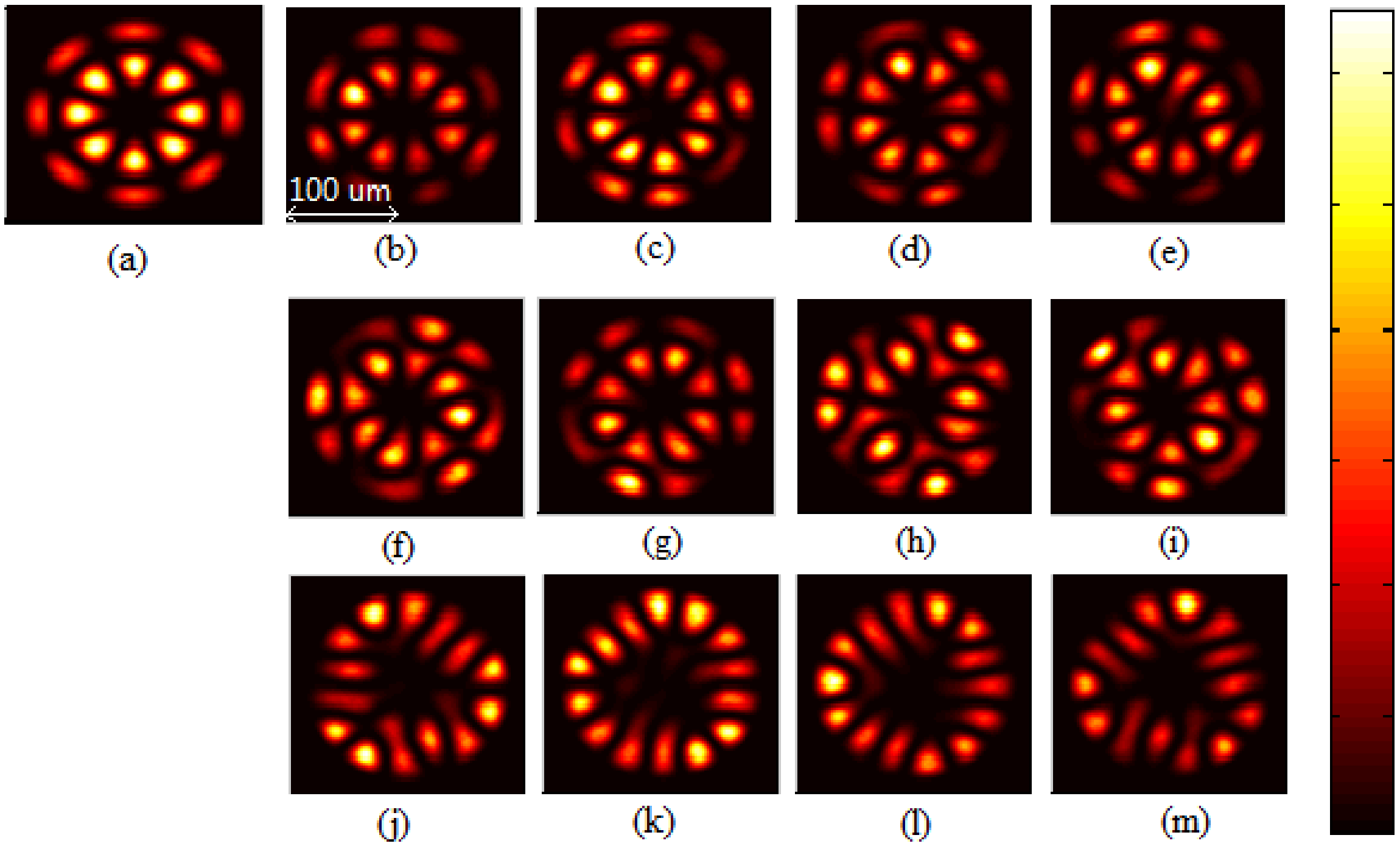

Fig. 13. Illustration of the spatial distribution for the 24$^{th}$ eigenmode of BECs in the lattice potential (Fig. 11). (a): For lattice potential without disorder, BECs are in shape of two distinct rings and they are evenly divided on the rings; (b)-(e): The first type of BECs distribution: two distinct rings, and BECs are not evenly distributed on the rings due to the localization effect; (f)-(i): The second type of BECs distribution: two rings that are interacting with each other; (j)-(m): The third type of BECs distribution: one ring that is merged by two.

If disorder exists, the shape of each realization is different, but they can be categorized by three types. The first type (Fig. 13-(b-e)) is a shape of two distinct rings, but BECs are not evenly divided on the rings due to the localization effect. The second type is a shape of two interacting rings (Fig. 13-(f-i)), i.e. the inner ring interacts with the outer ring. The third type is a shape of one ring that is merged by two (Fig. 13-(j-m)), namely, the inner and outer ring interact so much that they merge into one ring. For the low disorder level (1%), all the 120 simulations are the first type, therefore the shape of two distinct rings is the dominant shape. As the disorder level is higher (5%), we observe the first type for 24 times, the second type for 86 times, and the third type for 10 times, therefore the shape of two interacting rings dominates. For 30% disorder, we observe the first type for 6 times, the second type for 47 times, and the third type for 67 times, thus the shape of one ring that is merged by two is the dominant shape.

The nature of these statistically-stable shapes is unknown, and, probably, it is caused by Anderson localization in its formation stage (Fig. 13-(a) and Fig. 13-(b-e)). The two rings are almost uniformly divided in Fig. 13-(a) when there is no disorder. Small disorder of 1% leads to non-uniform distribution on the rings due to the localization effect. It can be seen that the inner ring will be shifted towards the outer ring for increased disorder, so that the inner and outer rings starts to interact and merge. We mostly see the shapes in Fig. 13-(a-e) for low disorder (≤1%), and the two rings are isolated. If the disorder level reaches 5%, one mostly observes that the two rings interact with each other (Fig. 13-(f-i)). The highest disorder in our test results in so much interactions of the inner and outer ring that they merge into one ring (Fig. 13-(j-m)). These facts make the further research interesting to find the edge-localized modes which can demonstrate the topological features towards the shape distortions [6],[16],[17],[30],[31].

## 5. Conclusion

In this paper, the eigenmodal iteration techniques have been improved and verified by published experimental data and the convergence criterion. The results obtained for the 2-D photonic lattices and non-interacting BECs captured in the 2-D and 3-D random speckle potentials show that the simulated localizations are presumably due to the Anderson effect taking into account its statistical stability of the modeled objects. Typically for the photonic waveguide case, our approach is more efficient in computation than the known split-step Fourier method. The main and high-order modes of 2-D BECs have been modeled, and the influence of the potential randomness on the level of their localization has been studied, including the edge-localization effect of these BECs. To avoid large diffusive component of BECs localized by speckles, the condensates have been proposed to be captured by a hologram forming a 3-D random lattice potential allowing increased localization of BEC due to better trapping of particles. The theoretically reached minimal size of the trapped BECs has been found close to a half of the lattice period, which is comparable with the particle wavelength for the given energy.

## 6. Acknowledgements

The first author (G. Ying) thanks the NTNU for his PhD grant. We greatly appreciate the access to the high-performing computing facilities granted by the Norwegian Metacenter for Computational Science (NOTUR-Project). The second author (G.A. Kouzaev) is grateful to the researchers from QSTAR at the European Laboratory of Nonlinear Sciences (LENS) (Florence, Italy) for initiating his interest in this matter during sabbatical stay there.

## References

[1] P.W. Anderson, Absence of diffusion in certain random lattices, Phys. Rev. 109 (1958) 1492-1505.

[2] M. Storzer, P. Gross, C.M. Aegerter, G. Maret, Observation of the critical regime near Anderson localization of light, Phys. Rev. Lett. 96 (2006) 063904(1-4).

[3] T. Schwartz, G. Bartal, S. Fishman, M. Segev, Transport and Anderson localization in disordered two-dimensional photonic lattices, Nature 446 (2007) 52-55.

[4] S. Karbasi, C.R. Mirr, P.G. Yarandi, R.J. Frazier, K.W. Koch, A. Mafi, Observation of transverse Anderson localization in an optical fiber, Opt. Lett. 37 (2012) 2304-2306.

[5] D. Wiersma, Disordered photonics, Nature Photon. 7 (2013) 188–196.

[6] M. Segev, Y. Silberberg, D.N. Christodoulides, Anderson localization of light, Nature Photon. 7 (2013) 197-204.

[7] S. Stützer, Y.V. Kartashov, V.A. Vysloukh, V.V. Konotop, S. Nolte, L. Torner, A. Szameit, Hybrid Bloch-Anderson localization of light, Opt. Lett. 38 (2013) 1488-1490.

[8] R. Dalichaouch, J.P. Armstrong, S. Schultz, P.M. Platzman, S.L. McCall, Microwave localization by two-dimensional random scattering, Nature 354 (1991) 53-55.

[9] A. Schelle, D. Delande, A. Buchleitner, Microwave-driven atoms: from Anderson localization to Einstein's photo effect, Phys. Rev. Lett. 102 (2009) 183001(1-4).

[10] S. Faez, A. Strybulevych, J.H. Page, A. Lagendijk, B.A. van Tiggelen, Observation of multifractality in Anderson localization of ultrasound, Phys. Rev. Lett. 103 (2009) 155703(1-4).

[11] J.E. Lye, L. Fallani, M. Modugno, D.S. Wiersma, C. Fort, M. Ingusciok, Bose-Einstein condensate in a random potential, Phys. Rev. Lett. 95 (2005) 070401(1-4).

[12] J. Billy, V. Josse1, Z. Zuo1, A. Bernard, B. Hambrecht, P. Lugan, D. Clement, L. Sanchez-Palencia, P. Bouyer, A. Aspect, Direct observation of Anderson localization of matter waves in a controlled disorder, Nature Lett. 453 (2008) 891-894.

[13] F. Jendrzejewski, A. Bernard, K. Mueller, P. Cheinet, V. Josse, M. Piraud, L. Pezze, L. Sanchez-Palencia, A. Aspect, P. Bouyer, Three-dimensional localization of ultracold atoms in an optical disordered potential, Nature Phys. 8 (2012) 398-403.

[14] L. Sanchez-Palencia, D. Clement, P. Lugan, P. Bouyer, G.V. Shlyapnikov, A. Aspect, Anderson localization of expanding Bose-Einstein condensates in random potentials, Phys. Rev. Lett. 98 (2007) 210401(1-4).

[15] A. Mackinnon, and B. Kramer, The scaling theory of electrons in disordered solids: additional numerical results, Z. Physik B: Condens. Matter, 53 (1983) 1-13.

[16] G.A. Kouzaev, Controlled localized eigenmodes in pseudo-random multilayer electron waveguides, Int. J. Mod. Phys. B 28 (2014) 135192(1-21).

[17] G.A. Kouzaev, Electronic control of edge modes in integer-Hall-effect 2D electron waveguides, http://arxiv.org/abs/1506.03294 (2016).

[18] R.H. Hardin, F.D. Tappert, Applications of split step Fourier method to the numerical solution of nonlinear and variable coefficient wave equations, SIAM Rev. Chron. 15 (1973) 423-423.

[19] F. Dalfovo, S. Giorgini, L.P. Pitaevskii, S. Stringari, Theory of Bose-Einstein condensation in trapped gases, Rev. Mod. Phys. 71 (1999) 463-512.

[20] G. Ying, G.A. Kouzaev. Solution of 2-D nonlinear Schrödinger equations by circuit simulators, in: G.A. Kouzaev, Applications of Advanced Electromagnetics, 2013, Springer, pp. 368-391.

[21] Y. Saad, Numerical Methods for Large Eigenvalue Problems, Manchester University Press, 1992.

[22] F.S. Cataliotti, L. Fallani, F. Ferlaino, C. Fort, P. Maddaloni, M. Inguscio, Dynamics of a trapped Bose–Einstein condensate in the presence of a one-dimensional optical lattice, J. Optics B: Quant. Semiclass. Optics 5 (2003) 17-22.

[23] G. Roati, C. D'Errico, L. Fallani, M.Fattori, C. Fort, M. Zaccanti, G. Modugno, M. Modugno, M. Inguscio, Anderson localization of a non-interacting Bose-Einstein condensate, Nature 453 (2008) 895-898.

[24] K. Murr, R. Nussmann, T. Puppe, M. Hijlkema, B. Weber, S.C. Webster, A. Kuhn, G. Rempe, Three-dimensional cavity cooling and trapping in an optical lattice, Phys. Rev. A 73 (2006) 063415(1-10).

[25] C. Dainty, Laser Speckle and Related Phenomena, Springer, 1984, pp. 9-68.

[26] S.S. Kondov, W.R. McGehee, J.J. Zirbel, B. DeMarco, Three-dimensional Anderson localization of ultracold matter. Science 334 (2011) 66-68.

[27] S. Equis, P. Jacquot, Simulation of speckle complex amplitude: advocating the linear model. Proc. SPIE6341, Speckle06: Speckles, From Grains to Flowers (2006) 634138(1-6).

[28] R. Roth, K. Burnett, Phase diagram of bosonic atoms in two-color superlattices, Phys. Rev. A 68 (2003) 023604(1-17).

[29] S. Suntsov, K.G. Makris, D.N. Christodoulides, G.I. Stegeman, A. Hache, R. Morandotti, H. Yang, G. Salamo, M. Sorel, Observation of discrete surface solitons, Phys. Rev. Lett., 96 (2006) 063901(1-4).

[30] Y.E. Kraus, Y. Lahini, Z. Ringel, M. Verbin, O. Zilberberg, Topological states and adiabatic pumping in quasicrystals, Phys. Rev. Lett., 109 (2012) 106402(1-5).

[31] M.C. Reichtsman, J.M. Zeuner, Y. Plotnik, Y. Lumer, D. Podolsky, F. Dreisow, S. Nolte, M. Segev, A. Szameit, Photonic Floquet topological insulators, Nature 496 (2013) 196-200.